\documentclass[prd,11pt,tightenlines,nofootinbib,superscriptaddress]{revtex4}
\usepackage{amsfonts,amssymb,amsthm,bbm,psfrag}

\usepackage{amsmath,amssymb,bbm}
\usepackage{amsfonts}

\newcommand{\N}{{\mathbb N}}
\newcommand{\R}{{\mathbb R}}

\newcommand{\cL}{{\mathcal L}}

\newcommand{\SU}{\mathrm{SU}}

\newcommand{\va}{\scriptscriptstyle}

\newcommand{\inter}{{\lrcorner}}

\newcommand{\be}{\begin{equation}}
\newcommand{\ee}{\end{equation}}
\newcommand{\beq}{\begin{eqnarray}}
\newcommand{\eeq}{\end{eqnarray}}
\newcommand{\bea}{\begin{eqnarray}}
\newcommand{\eea}{\end{eqnarray}}

\newcommand{\f}{\frac}

\newcommand{\rd}{\mathrm{d}}

\newcommand{\bz}{\overline{z}}

\usepackage{latexsym}
\usepackage{amsmath,amsfonts}
\usepackage{amsbsy}
\usepackage{mathrsfs}
\usepackage{color}

\usepackage{enumerate}

\usepackage{amsmath,amssymb,calc,amsfonts}
\usepackage{latexsym}

\usepackage{graphicx,calc,epsfig}
\def\ut#1{\rlap{\lower1ex\hbox{$\sim$}}#1{}}
\newcommand{\ba}{\nopagebreak[3]\begin{eqnarray}}
\newcommand{\ea}{\end{eqnarray}}
\DeclareFontFamily{U}{rsfs}{}         
\DeclareFontShape{U}{rsfs}{m}{n}{<5> rsfs5 <6><7> rsfs7          %
  <8><9><10><10.95><12><14.4><17.28><20.74><24.88> rsfs10}{}     %
\DeclareMathAlphabet{\mathfs}{U}{rsfs}{m}{n}                     %
\newcommand{\e}{\mathbf  e}\newcommand{\n}{\nonumber}
\newcommand{\ez}{ {e^i_{z}}}
\newcommand{\ezb}{ {e^i_{\bar z}}}

\def\pb#1{\rlap{\lower1.5ex\hbox{$\longleftarrow$}}{#1}}
\def\dpb#1{\rlap{\lower1.5ex\hbox{$\Longleftarrow$}}{#1}}
\def\spb#1{\rlap{\lower1.5ex\hbox{$\leftarrow$}}{#1}}
\def\sdpb#1{\rlap{\lower1.5ex\hbox{$\Leftarrow$}}{#1}}
\newcommand{\red}[1]{{\color{red} #1}}

\definecolor{blue}{rgb}{0,0,1}
\definecolor{green}{rgb}{0,1,0}
\definecolor{red}{rgb}{1,0,0}
\definecolor{vio}{rgb}{1,0,1}
\definecolor{ama}{rgb}{1,1,0}

\begin{document}

\title{Quantum gravity at the corner}

\author{Laurent Freidel}
\affiliation{ Perimeter Institute for Theoretical Physics,
Waterloo, N2L-2Y5, Ontario, Canada.}
\author{Alejandro Perez}
\affiliation{
    Aix Marseille Universit\'e, CNRS, CPT, UMR 7332, 13288 Marseille, and
    Universit\'e de Toulon, CNRS, CPT, UMR 7332, 83957 La Garde, France.
}

\begin{abstract}
We investigate the quantum geometry of $2d$  surface $S$ bounding the Cauchy slices of 4d gravitational system.
We investigate in detail and for the first time the symplectic current that naturally arises boundary term in the first order formulation of general relativity in terms of the Ashtekar-Barbero connection.
This current is proportional to the simplest quadratic form constructed out of the  triad field, pulled back on  $S$.
We show that the would-be-gauge degrees of freedom---arising from $SU(2)$ gauge transformations plus diffeomorphisms tangent to the boundary,
are entirely described by the boundary $2$-dimensional symplectic
form and give rise to a representation at each point of $S$  of 
$SL(2,\mathbb{R}) \times SU(2)$.
Independently of the  connection with gravity, this system is very simple and 
rich  at the quantum level with possible connections with conformal field theory in 2d. A direct application of the quantum theory is 
modelling of the black horizons in quantum gravity.

\end{abstract}

\maketitle



\section{Introduction}

In the construction of black hole models in loop quantum gravity \cite{G.:2015sda, BarberoG.:2012ae} via the so-called isolated horizon boundary condition \cite{Ashtekar:2004cn}
the  boundary would-be-gauge degrees of freedom are described by a  Chern-Simons theory living 
on the black hole horizon \cite{ABK, Engle:2009vc, Engle:2010kt,Perez:2010pq}.  The appearance of the specific Chern-Simons boundary dynamics is usually argued to be due to restrictions on the set of boundary conditions adapted to isolated horizons.
What we realise here is that the appearance of a boundary dynamical theory and the appearance of a boundary symplectic structure is  not specific of black holes and arises naturally in the most general situation \cite{Freidel:2014qya}. 
As we explain, the  general boundary dynamics can be understood in terms of a Chern-Simons theory. But this Chern-Simons theory doesn't need the introduction of auxiliary fields. Remarkably it can be expressed very simply in terms  of the pull back of the triad frame field on the boundary, while the pull back of the spin connection acts as a Lagrange multiplier for the boundary diffeomorphisms. 
The Boundary symplectic structure is remarkably simple, it reads 
\be
\Theta =\frac1{2\gamma} \int_{\partial \Sigma} \delta e^i \wedge \delta e_i,
\ee
where $\gamma$ is the Immirzi parameter and $ e^i$ the triad field pull back on the 2d boundary of the slice $\Sigma$.
This  remarkably simple and natural boundary structure  constitute one of the central building blocks of first order gravity theory projected on any  corner sphere.
In this paper we provide the detailed proof that such a symplectic structure allow a complete Hamiltonian  description of the boundary gauge diffeomorphisms transformations. These  shows that these would-be-gauge degrees of freedom exhausts the set of boundary degrees of freedom.
 This symplectic structure first made its appearance in \cite{Engle:2010kt}, (see also \cite{Bodendorfer:2013sja} for a discussion in higher dimension) but its central importance was not emphasized and it was not studied in full generality.
 At first sight such theory would seem harder to quantize as the standard techniques developed for the background independent quantization of connections cannot be directly applied. 
However,  quantisation is made possible by the choice of a complex structure on the $2$-dimensional boundary associated with  fiducial coordinates.  This leads to expressing the triad in terms of harmonic oscillators associated to point defects (punctures) on the boundary. The unrestricted Hilbert space is much larger than the one found for quantum isolated horizons as expected from the fact that no classical symmetry reduction  on the geometry of the boundary has been imposed. We show that the representations of the geometric observables can be constrained in a simple way in order to recover the usual accounts of black hole entropy in the literature.   

The paper is organized as follows. In the following section we describe the geometric context in which the $2$-dimensional model 
we analyze is natural. We also show how in the situations where $SU(2)$ gauge transformations and bulk diffeomorphism that are tangent to the boundary are gauge symmetries of
gravity. In section \ref{sym} we analyse the boundary symplectic structure and define the associated three dimensional theory encoding the entire dynamics of the would-be-gauge degrees of freedom is controlled by our $2+1$-dimensional system. In Section \ref{II} we quantize the system and
interpret the states in terms of the underlying complex structure. We close the paper with some concluding remarks in relation to the applicability of our results for the computation of black hole entropy in Section \ref{III}.

\begin{figure}[h] \centerline{\hspace{0.5cm} \(
\begin{array}{c}
\psfrag{a}{$M$}
\psfrag{b}{$\Sigma$}
\psfrag{c}{$\Sigma'$}
\psfrag{d}{$\partial\Sigma$}
\psfrag{e}{$\partial\Sigma$}
\includegraphics[width=7cm]{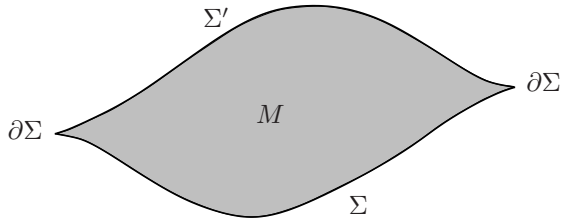}
\end{array}
\)}
\caption{Spacetime region obtained from the time flow that is allowed in our analysis. Lapse and shift are constrained on the corners $\partial \Sigma$ in order to preserve the boundary fixed up to tangent diffeomorphisms and gauge transformations.} \label{corners}
\end{figure}

\section{The origin of the 2d symplectic structure}
Starting from the first order formulation of gravity whose action is
\be S[e,\omega]=\int_M \epsilon_{IJKL} e^I\wedge e^K\wedge F^{KL}(\omega),\ee
introducing a foliation of $M$ in terms of Cauchy surfaces $\Sigma$,  and using the time gauge $e^0=n$ where $n$ is the co-normal to $\Sigma$, the 
canonical symplectic structure of gravity takes the form
\be
\Omega_C= \int_{\Sigma} \delta {K}^{i} \wedge \delta \Sigma_{i},
\ee
Where $K^i\equiv \omega^{0i}$ is the extrinsic curvature one-form and 
$\Sigma=\tfrac12 \epsilon_{ijk} e^j\wedge e^k$ is the flux two form.
Here and in the following $\delta$ denotes the differential on field space, in particular as a differential it anti-commute with itself and its square $\delta^2=0$ vanishes.
It should not be confused with $\rd$ which denotes the differential on space.
The symplectic form in Ashtekar-Barbero variables is given by 
\be
\Omega \equiv \frac1{\gamma} \int_{\Sigma } \delta A^{i}\wedge \delta \Sigma_{i}
\ee
where $A^{i}$ is the SU$(2)$ connection, which can be expressed as $ A^{i}=\Gamma^{i}+ \gamma K^{i}$,
in terms of the spin connection $\Gamma^{i}\equiv \frac12 \epsilon^{i}{}_{jk}\omega^{jk}$ and the 
extrinsinc curvature tensor  $ K^{i}\equiv \omega^{0i}$ with $\omega^{IJ}$ the Lorentz connection.
In the absence of boundaries one has that $\Omega =\Omega_C$ and this is the celebrated result that allow to see the previous connection as a canonical transformation from the original vector variables \cite{Barbero:1994ap, Immirzi:1996dr}.
In the presence of a boundary $\partial\Sigma\not=0$ (see figure) one has that \be\Omega_C=\Omega +\Theta, \ee 
where $\Theta$ is
boundary symplectic structure  given by \cite{Engle:2010kt}
\be
\Theta \equiv \frac1{2 \gamma} \int_{\partial \Sigma} \delta e^{i} \wedge \delta e_{i},
\ee
as it  follows from \beq
 \Theta = \frac1{2 \gamma}\int_{\Sigma}\rd(\delta e_{i} \wedge \delta e^{i}) =
 - \f1\gamma \int_{\Sigma} \delta {\Gamma}^{i} \wedge \delta [e,e]_{i}.\
\eeq
and from the identification 
\be\label{relation}
\Sigma_i = \frac12 [e,e]_i
\ee
which is valid at the boundary.

\subsection{Symmetries}
In this section we analyse the transformation property of the symplectic form 
$\Omega +\Theta$ under two types of transformations: SU(2) gauge transformations labelled by $\alpha \in \mathfrak{su}(2)$, and 
spatial diffeomorphism labelled by a vector field $\xi$.
Our variables are the bulk variables $(\Sigma_i, A^i)$: a Lie algebra valued two-form and an SU$(2)$ connection on $\Sigma$;  and the boundary variables $e_i$, which is a Lie algebra valued one-form on $\partial \Sigma$.
We initially treat these variables as independent variables. As we will see, the gauge symmetry will restore the relationship (\ref{relation}) at the boundary.

The gauge transformations are labelled by an SU$(2)$ lie algebra element $\alpha^i$ and are defined to be 
\be
\delta_{\alpha} A \equiv-\rd_{A} \alpha,\qquad 
\delta_{\alpha}\Sigma \equiv [\alpha,\Sigma],
\qquad
\delta_\alpha e_i \equiv  [\alpha, e].
\ee
Infinitesimal diffeomorphisms are labelled by a vector field $\xi$ and generated by the Lie derivative 
\be
{\cal L}_\xi \equiv \rd i_\xi + i_\xi \rd,
\ee
where $i_{\xi} T_{bc\cdots d}\equiv \xi^a T_{abc\cdots d} + \xi^aT_{b ac\cdots d} +\cdots, $ is the inner contraction for an arbitrary tensor $T_{abc\cdots d}$.
This Lie derivative has the disadvantage of not preserving the 
SU$(2)$ covariance when acting  on SU$(2)$ tensors since it doesn't commute with gauge transformations $[{\cal L}_\xi, \delta_\alpha] \neq0$.
For that reason it is more natural to work with a gauge invariant Lie derivative
denoted $L_\xi$ which   preserve the covariance under gauge transformations : $[L_\xi ,\delta_\alpha]=0$. This covariant Lie derivative acts on SU$(2)$ tensors like $e_i$ or $\Sigma_i$ or $F^i(A)$  as 
\be\label{Le}
L_\xi\equiv \rd_A i_\xi + i_\xi \rd_A,
\ee
but it acts differently on the gauge connection\footnote{It is easy to check that due to the Bianchi identity, the definitions (\ref{LA}) and  (\ref{Le}) are equivalent  for $F(A)$.} since 
\be\label{LA}
L_\xi A \equiv i_\xi F(A).
\ee
This covariant Lie derivative restricts to the usual Lie derivative for SU$(2)$ scalars.
On SU$(2)$ tensors the covariant and usual Lie derivative are equivalent up to gauge transformations,  the relation is simply
\be
L_\xi =  {\cal L}_\xi + \delta_{i_\xi A}.
\ee
In the following we uses $L_\xi$ as the generator of covariant diffeomorphisms.
$\xi$ is a vector field on $\Sigma$ which is assumed to be tangent to $\partial \Sigma$.
Therefore $\xi$ labels an infinitesimal diffeomorphism of $\Sigma$ which do not move the boundary.

%
\subsection{Hamiltonian generators}

The goal of this section is to show that the  Hamiltonian generators 
of  covariant diffeomorphisms $L_\xi$ and gauge  symmetry $\delta_\alpha$  are given by 
\beq
H_{\xi}\equiv  \int_{\Sigma}  i_{\xi}F \wedge \Sigma + \frac12\int_{\partial\Sigma} L_\xi e^i \wedge e_i,
\qquad 
G_{\alpha}\equiv
- \int_{\Sigma} \rd_{A}\alpha^i\wedge \Sigma_i +\frac{1}{2} \int_{\partial \Sigma}\alpha_i [e,e]^i.
\eeq
We start by computing  the variation of the gauge Hamiltonian.
\beq\label{gagy}\delta G_{\alpha} 
&=& \n-\int_{\Sigma}\rd_{A} \alpha_{i} \wedge \delta \Sigma^{i} - 
\int_{\Sigma} \delta A^{i}\wedge [\alpha, \Sigma] _{i}{+ \int_{\partial\Sigma} [\alpha, e]^{i}\wedge \delta e_{i}  }\\
&=&\n\int_{\Sigma}\delta_{\alpha} A_{i} \wedge \delta \Sigma^{i} - \int_{\Sigma} \delta A_{i}\wedge \delta_{\alpha} \Sigma^{i}{+ \int_{\partial\Sigma} \delta_\alpha e^{i}\wedge \delta e_{i}  }\\
&=& \gamma\, \delta_{\alpha} \inter (\Omega+{\Theta})
\eeq
where $\delta_{\alpha} \inter \Omega$ denotes the interior product of the field variation
$\delta_{\alpha}$ with the field two form $\Omega+\Theta$. This shows that  $G_\alpha$ is the Hamiltonian generating $SU(2)$ gauge transformations. This generator is the sum of  a bulk and a boundary terms. The bulk constraint imposes the Gauss law while the boundary constraints imposes a soldering of the boundary degree of freedom to the bulk degree of freedom. Integrating by part we can write
\be
G_\alpha = \int_{\Sigma} \alpha^i\wedge \rd_{A} \Sigma_i + \int_{\partial \Sigma}\alpha_i \left( \frac{1}{2}[e,e]^i -\Sigma^i\right)
\ee
 In short $G_\alpha=0$ means that
\be
\rd_A \Sigma_i=0,\qquad \Sigma_i=\frac12[e,e]_i.
\ee  The first condition is the usual  Gauss law. The second one is a first class boundary constraint simply demands that the induced area density from the bulk and the intrinsic one match\footnote{In the Chern-Simons description of the boundary degrees of freedom that is used in applications to isolated horizons the fusion conditions between the boundary induced connection and $\Sigma$ involves components of the Weyl curvature \cite{Ashtekar:2004nd, Beetle:2010rd}. This requires the definition of a new boundary connection that is related to the original one in a non-trivial fashion making the final structure geometrically obscure. As we see here the Bulk boundary connection is extremely natural}. 

It is convenient to introduce the boundary variation and hamiltonian:
\be
\delta_{\alpha} e^{i} \equiv [\alpha, e]^{i},\qquad g_{\alpha} \equiv \int_{\partial\Sigma}  \f12 \alpha_{i} [e, e]^{i}.
\ee
$g_\alpha$ is the generator of boundary variations. It doesn't act on the bulk fields but is the Hamiltonian for boundary rotations:
\bea
\delta g_{\alpha} = \int_{\partial \Sigma} [\alpha, e]^{i}\wedge \delta e_{i} = 
\gamma 
\delta_{\alpha}\inter \Theta.
\eea

We now do the same computation for the diffeomorphism variation.
This computation is more involved and in order to do it we separate the 
bulk and boundary variations. 
We start with 
\beq
\delta \left(\int_{ \Sigma} i_{\xi } F_{i} \wedge \Sigma^{i}\right)
&=&\n \int_{ \Sigma} i_{\xi } \rd_{A} \delta A_{i} \wedge \Sigma^{i}   + \int_{\Sigma} i_{\xi }F^{i}\wedge \delta \Sigma_{i} 
\\
&=&  \n-\int_{ \Sigma}  \rd_{A} \delta A_{i} \wedge i_{\xi } \Sigma^{i}   + \int_{\Sigma} i_{\xi }F^{i}\wedge \delta \Sigma_{i} \\
 &=& \n - \int_{\partial \Sigma} \delta A_{i} \wedge i_{\xi}\Sigma^{i} + \int_{\Sigma} i_{\xi }F^{i}\wedge \delta \Sigma_{i}  
  -\int_{ \Sigma}   \delta A_{i} \wedge \rd_{A} (i_{\xi } \Sigma^{i}) \\
  &=& \n \int_{\partial \Sigma} i_{\xi}\left(\delta A_{i} \wedge \Sigma^{i} \right)  + \int_{\Sigma} i_{\xi }F^{i}\wedge \delta \Sigma_{i} 
  -\int_{ \Sigma}   \delta A_{i} \wedge L_\xi \Sigma^{i}
  -\int_{ \Sigma}   \delta A_{i} \wedge i_{\xi }( \rd_{A} \Sigma^{i})
  -  \int_{\partial \Sigma} (i_{\xi}\delta A_{i}) \wedge \Sigma^{i}\\
&=& \n \int_{\partial \Sigma} i_{\xi}\left(\delta A_{i} \wedge \Sigma^{i} \right) +\gamma \delta_{\xi} \inter \Omega + G_{(i_{\xi}\delta A)}  - g_{(i_{\xi}\delta A)}\\
&=&  \gamma L_{\xi} \inter \Omega + G_{(i_{\xi}\delta A)}  - g_{(i_{\xi}\delta A)}\label{Bulkv}
\eeq
where on the last line we have used that $\xi|_{\partial\Sigma}$  is a vector tangent to 
$\partial  \Sigma$. 
We can now focus on the boundary term variation.
We define the boundary hamiltonian 
\be
h_\xi \equiv \frac12 \int_{\partial \Sigma} L_{\xi} e_i \wedge e^i.
\ee
Its variation is given by 
\bea
\delta h_\xi &=& \n\frac12 
\int_{\partial \Sigma}  [(i_\xi\delta A), e_i] \wedge e^i + \frac12
\int_{\partial \Sigma} L_{\xi} \delta e_i \wedge e^i
+\frac12 \int_{\partial \Sigma} L_{\xi} e_i \wedge \delta e^i \\
\n 
&=& 
\n
 \int_{\partial \Sigma} (i_\xi \delta A_i) \wedge \frac12 [e, e]^i
+ 
\frac12 \int_{\partial \Sigma} L_{\xi} \left(\delta e_i \wedge e^i\right)
+ \int_{\partial \Sigma} L_{\xi} e_i \wedge \delta e^i \\
&=& 
\n
\frac12\int_{\partial \Sigma} i_\xi\rd \left( \delta e_i \wedge e^i\right)
+g_{i_{\xi}\delta A} + \gamma L_{\xi}   \inter \Theta\\
&=& \gamma  L_{\xi}  \inter \Theta + g_{i_{\xi}\delta A},\label{boundv}
\eea
where we have assumed again that $\xi$ is a vector tangent to the boundary.
Taking the sum of (\ref{Bulkv}) and (\ref{boundv}) gives 
\be\label{Hxi}
\delta H_\xi = \gamma L_\xi  \inter (\Omega+\Theta ) +  G_{i_{\xi}\delta A}.
\ee
We can also use the gauge variation, computed already in (\ref{gagy}), to establish that 
\be
\delta G_{i_{\xi}A} ={ \gamma \delta_{i_{\xi}A} \inter (\Omega+\Theta)}
+  G_{i_{\xi}\delta A}.
\ee
This implies that if one  introduces the generator of (non-covariant) diffeomorphism 
$D_{\xi } \equiv H_{\xi } - G_{i_{\xi}A}$. Taking the difference of the previous equalities one obtain that 
\be
\delta D_\xi = \gamma {\cal L}_\xi  \inter (\Omega+\Theta ).
\ee
Imposing the covariant diffeomorphism constraints implies that we impose a bulk and a boundary constraints given by 
\be
(i_\xi F_i)\wedge \Sigma^i =0, \qquad h_\xi=0.
\ee
The boundary constraint can be expressed more explicitly as 
\bea
h_\xi &=&\frac12 \int_{\partial \Sigma}  \rd_\Gamma (i_\xi e_i) \wedge e^i + 
\frac{\gamma}2 \int_{\partial \Sigma} 
(i_\xi K_i) \wedge [e,e]^i\\
&=& \frac{\gamma}2 \int_{\partial \Sigma} 
(i_\xi K_i) \wedge [e,e]^i = \gamma g_{i_\xi K} .
\eea
In order to understand the meaning of the condition $h_\xi=0$ for all $\xi$ tangent to $\partial \Sigma$ we now study its geometrical meaning which 
  follows from the following analysis:
 We can write $K^i_a=\alpha^{ij} e_{ai}$ and the imposition of the Gauss Law implies that $K^{ij}$ is a symmetric internal tensor.
  The extrinsic curvature can be written as   $K_{ab}=K_{ij}e^i_ae^j_b$. 
  We introduce $N_a$ a spatial unit vector,  to be the normal of $\partial \Sigma$ within $\Sigma$ and we go  to the gauge where $e^3_a=N_a$. The condition $h_{\xi}=0$ implies that 
$\alpha^{3A}=0$ for $A=1,2$ which means that the second fundamental form is 
$K_{ab}=\alpha_{AB} e^A_ae^B_b+\alpha_{33}e^3_ae^3_a$ or simply that
 \be
 K_{ab}= k_{ab}+\alpha_{33} N_aN_b,
\ee 
where $k_{ab}$ is a symmetric tensor tangent to $\partial \Sigma$, i.e. $k_{ab}N^a=0$.
$k_{ab} $ is the 2d extrinsic curvature of $\partial \Sigma$ as embedded in $\Sigma$ 
The 3d extrinsic curvature $K_{ab}$ can be written as $K_{ab}=\frac{1}{3}\theta q_{ab} +\sigma_{ab}$, i.e., into its trace part (the expansion) $\theta$ and traceless part (the shear) $\sigma_{ab}$.
The previous expression implies that the shear
 \be
 \sigma_{ab}=\sigma^{(2)}_{ab}+\sigma_{33} N_aN_b,
\ee 
 which means that $N_a$ is one of the principal axis of the shear while the other two are tangent to $\partial \Sigma$.
 The geometric interpretation is now clear: an infinitesimal spherical ball around a point at $\partial \Sigma$ when propagated along the timelike 
 geodesics normal to $\Sigma$ is allowed to expand and deform along directions which are either normal or tangent to $\partial \Sigma$. Deformation in an alternative direction is precluded by $h_{\xi}=0$.  { This can be interpreted as a condition of {\em non-rotation} for the boundary $\partial \Sigma$. 
In the case when an axisymmetry Killing field $\xi$ tangent to $\partial \Sigma$ is available then $h_{\xi}$ is exactly the Komar angular momentum\footnote{When available, the Komar angular momentum is given by \be J=\frac{1}{8\pi} \int_{\partial \Sigma} \epsilon_{abcd} \nabla^{a}\xi^b=-\frac{1}{4\pi} \int_{\partial \Sigma} (N^a n^b \nabla_{a}\xi_b) \epsilon_{ab}=\frac{1}{4\pi} \int_{\partial \Sigma} \xi^b N^a (\nabla_{a}n_b)  \epsilon_{cd} = \frac{1}{4\pi} \int_{\partial \Sigma}  \sqrt{q} K_{a3}\xi^a,
\ee
where $n^a$ is the normal to $\Sigma$, $N^a$ is the normal to $\partial\Sigma$, $\epsilon_{ab}$ is the volume form of $\partial\Sigma$, and $\epsilon_{abcd}=-12 N_{[a}n_b\epsilon_{ab]}$ is the spacetime volume form. The last expression is obtained using $N^a n^b \nabla_{a}\xi_b=\nabla_{a}(N^a n^b \xi_b)-\nabla_{a}(N^a n^b) \xi_b$.  The last expression is proportional to $h_{\xi}$ in the normal gauge.}
.}

\begin{figure}[h] \centerline{\hspace{0.5cm} \(
\begin{array}{c}
\psfrag{a}{$\Sigma$}
\psfrag{b}{$\partial\Sigma$}
\includegraphics[width=14cm]{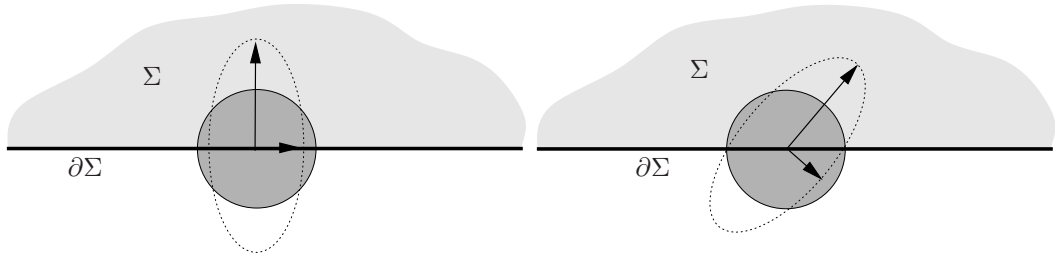}
\end{array}
\)}
\caption{Deformation of a ball of geodesics normal to $\Sigma$. On the left panel $h_{\xi}=0$:  the principal axis of the shear are tangent to $\partial \Sigma$ and normal to $\partial \Sigma$. On the right panel $h_{\xi}\not=0$, the boundary ``moves".} \label{hxi}
\end{figure}

From (\ref{Hxi}) the poisson bracket of two Hamiltonian is therefore given by
\be
\{H_{\xi},H_{\xi'}\} = L_{\xi}\inter( L_{\xi'}\inter( \Omega +\Theta))
= L_{\xi}H_{\xi'} = H_{[\xi,\xi']}.
\ee
Thus on-shell (i-e when $i_\xi F^i \wedge \Sigma_i=0$)  the commutation relation of the angular momenta  simply gives a representation of the  2-dimensional diffeomorphism algebra:
\be
\{ h_{\xi},h_{\xi'}\} \simeq h_{[\xi,\xi']}.
\ee
 where $[\xi,\xi']$ is the Lie-bracket of the two vector fields. Let us finally remember that the boundary of generator of diffeomorphism is given by $ d_\xi = h_\xi - g_{i_\xi A}$ and explicitly expressed as 
 \be
 d_\xi = \frac12 \int_{\partial \Sigma} {\cal L}_\xi e_i \wedge e^i.
 \ee

 Non-static boundaries for which $h_{\xi}\not=0$ are physically very interesting (an important example is the Kerr black hole horizon when treated as a boundary). The presence of angular momentum however makes the question of diffeomorphims invariance more subtle and this introduces additional complication when one aims at the quantisation of the boundary would-be-gauge degrees of freedom. For an exploration of the quantisation of a non-static boundary see \cite{Frodden:2012en}. For that reason in what follows we will restrict to the static case $h_{\xi}=0$.




\section{Boundary symplectic structure}\label{sym}

The previous section shows that for the set of variation generated by gauge and  diffeomorphism the 
bulk symplectic structure is equivalent to the boundary symplectic structure
\be\label{sympl}
\Theta\equiv \frac{1}{2\gamma} \int_{H} \delta e_{i} \wedge \delta e^{i}.
\ee
This symplectic structure control the ``would be gauge'' degrees of freedom.
The remarkable property of this symplectic structure is that it leads to a non 
commutative flux algebra.
Indeed, defining for $S\subset H$
\be
X_{\alpha}(S) \equiv \int_{S} \Sigma_{i} \alpha^{i}
\ee
we have 
\be
\{X_{\alpha}(S),X_{\beta}(S')\}=\gamma X_{[\alpha,\beta]}(S\cup S').
\ee
In terms of the components $e_{A}^{i}$ the Poisson structure reads
\be\label{piosson}
\{e_{A}^{i}(x),e_{B}^{j}(x')\} = \gamma\epsilon_{AB} \delta^{ij} \delta^{2}(x-x')
\ee
Note that if we define some integrated version of the frame field along curves $C$: $e_{A}^{i}(C) =\int_{C} e^{i}(x) $ we obtain the loop algebra
\be
\{e_{A}^{i}(C),e_{B}^{j}(C')\} =\gamma N_{C\cap C'} \epsilon_{AB} \delta^{ij}
\ee
where $N_{C\cap C'}$ is the number of intersection of $C$ with $C'$ with positive orientation minus the 
number of intersection with negative orientation.

\subsection{The associated boundary $2+1$ dynamical theory}\label{eft}

Here we write a $2+1$ dynamical theory from which the $2d$ boundary symplectic structure (\ref{sympl}) arises in the canonical analysis.  In addition the constraint structure of the theory is compatible with the gauge symmetries expected to be relevant for the boundary degrees of freedom in view of eventually coupling them to the bulk quantum gravitational degrees of freedom of the ambient $3d$ quantum geometry.

Consider the $2+1$ action on $\partial\Sigma \times \mathbb{R}$
\be\label{effaction}
S[\bar{e}^i,\bar{\omega}^i]=-\frac{1}{\gamma}\int \bar{e}_i\wedge(d\bar{e}^i+ \epsilon_{ijk}\, \bar{\omega}^{j}\wedge \bar{e}^k).
\ee
First order variations of this action yield the symplectic structure (\ref{sympl}) and the equations of motion telling us that $\bar{\omega}_i$ is simply a Lagrange multiplier imposing $\bar{e}^i\wedge \bar{e}^j\,\epsilon_{ijk}=0$ and that $d_{\omega}\bar{e}^i=0$. 
There are non trivial solutions corresponding to degenerate triads. The degeneracy condition demands that 
$e_a^i $ is a matrix of rank one. 

The previous  action is the analog of the Chern-Simons action in the effective treatments of \cite{ABK,Engle:2010kt, Engle:2009vc}. However, unlike the latter the present one does have local degrees of freedom and this will explicitly show up in the quantisation. The present dynamical framework is therefore more general as expected from the fact that, in contrast to the approach leading to the Chern-Simons formulation,  we have not imposed any symmetry restriction on the boundary geometry.

The canonical analysis of (\ref{effaction}) yields the Poisson brackets (\ref{piosson}). Taking a $2+1$ decomposition $ \bar{e}_i = \beta_i \rd t + {e}_i$ 
and  $\bar{\omega}^i =\alpha^i \rd t + {\omega}^i$, where the barred form are 2 dimensional, we find that 
$S =\int \rd t L $ with
\be
L=
\frac1{2\gamma}\int_{\partial \Sigma} 
{e}^i\wedge  \partial_t {e}_i - 
\frac1{\gamma}\int_{\partial \Sigma} \alpha_i [e,e]^i + 
\beta_i  \rd_{{\omega}} {e}^i+\gamma_i \Pi^i_{\omega},
\ee
where $\Pi_{\omega}$ is the momentum conjugate to $\omega$.
The Hamiltonian is a linear combination of  primary constraints:\bea\label{45}
g(\alpha) = \int_{\partial \Sigma} \alpha_i [e,e]^i,
\qquad 
d(\beta) = \int_{\partial \Sigma} \beta^i \rd_{\omega} e_i\qquad 
\Pi(\gamma) = \int_{\partial \Sigma} \gamma_i \Pi^i_{\omega}.
\eea
The first one is the Gauss law that implies that $e_i$ is degenerate while the second one implies that $e^i$ is $\omega$-closed. The requirement that $\Pi_{\omega}$ is preserved by time evolution implies that
\be\label{this}
[\beta, e]^i=0.
\ee
This condition reduces the constraint system to the following first class system
\bea\label{45}
g(\alpha) = \int_{\partial \Sigma} \alpha_i [e,e]^i,
\qquad 
d(\beta) = \int_{\partial \Sigma} \beta^i \rd e_i\qquad .
\eea
Equation (\ref{this}) determines the Lagrange multiplier $\beta^i$.
If $e^i_a$ is of maximal rank $2$ it implies that $\beta=0$. 
When $e^i_a$ is degenerate of rank $1$ this equation is solved by  the choice of Lagrange multiplier $\beta^i=v^b e_b^i$ 
which when replaced in $d(\beta)$ gives
\bea\label{ww}
 d(\beta) =
  \int_{\partial \Sigma} (i_\beta e^i)  \rd e_i 
  = \frac12  \int_{\partial \Sigma} e_i \wedge {\cal L}_\beta(e^i), 
  \eea
  which reduces to the diffeomorphism constraints.
Therefore $d(v)$  is equivalent to the diffeomorphism constraints when $e_i$ is invertible. When $e_i$ is not invertible, it is more restrictive.


A naive counting of degrees of freedom would lead to the incorrect conclusion that this theory is topological.
However, further scrutiny shows, as we have seen that the field theory has local degrees of freedom corresponding to degenerate metric configurations.   In addition to these, the theory can acquire additional degrees of freedom if appropriately coupled with external charges which take the form of defects to the gauge constraints. 

For instance an external electric field can couple to the Chern-simons theory via 
\be
S_{\mathrm int} =\frac1{\gamma} \int \bar{\omega}^i\wedge \Sigma_i.
\ee 
This coupling is gauge invariant if the flux $\Sigma$ satisfies the Gauss law
$\rd_{\bar{\omega}}\Sigma =0$.
The addition of this term  gives the equation of motion
\be
[e,e]_i =\Sigma_i.
\ee
This will become apparent in the treatment in the following section.

\section{Quantisation: the discrete representation }\label{II}
We now study the quantisation of the Poisson algebra (\ref{piosson}).
In order to do so and since this algebra is ultralocal, we first perform a discretisation of the $2$d sphere in terms of a system of curves.
In order to define the discretisation we 
{ start from a conformal structure, this singles out an $dx$ and $dy$ ($dz, d\bar z$). We now introduce a set of paths 
$\{(L_x,L_y)\}$ and define $e^i(L_x)\equiv\int_{\va L_x} \!\!e^i$ and $e^i(L_y)\equiv\int_{\va L_y} \!\!e^i$ at every point of square lattice defined by the conformal structure (see figure \ref{figure}). It follows that
\be [e^i(L_x),e^j(L_y) ]=i \gamma \delta^{ij} \label{una}\ee
\be\Sigma_i=\epsilon_{ijk}(e^j(L_x)e^k(L_y)-e^j(L_x)e^k(L_y))\label{otra} \ee
} It will be convenient from now on to use an index notation $A,B$ instead of the explicit mentioning of $L_x$ and $L_y$. In addition we absorve the factor $\gamma$ defining
\be
e_A^i\equiv \frac {1}{\sqrt{\gamma}} e^i(L_A)
\ee  In this notation the the finite dimensional algebra smeared frame fields becomes
\be\label{53}
[ e_{A}^{i},e_{B}^{j}] = i  \epsilon_{AB} \delta^{ij}
\ee
Given the frame field we can define the flux and the metric
\be
\Sigma_{i} \equiv \frac12 \epsilon_{ijk} e^{j}_{A}e^{k}_{B} \epsilon^{AB},\qquad g_{AB} \equiv  e^{i}_{A}e^{i}_{B}.
\ee
These satisfy the algebra
\beq \label{alg}
[\Sigma^{i},\Sigma^{j}] &=& i \epsilon^{ijk}\Sigma_{k},\qquad [\Sigma^{i},g_{AB}]=0,
\\
{[}g_{AB},g_{A'B'}]&=& i(g_{AA'}\epsilon_{BB'}+g_{AB'}\epsilon_{BA'}+g_{BA'}\epsilon_{AB'}+g_{BB'}\epsilon_{AA'}).
\eeq
Moreover it is important  to note that 
$$ \det(g) = \Sigma_{i}\Sigma^{i},$$ and is therefore a casimir of this algebra.
One sees that $\Sigma^{i}$ capture the gauge degrees of freedom, $g_{AB}$ the metric degrees of freedom while 
the conformal degree of freedom is shared by both due to the previous relation.

We chose complex coordinates $z,\bar{z}$ on $H$ where $z=(x+iy)/\sqrt{2}$. 
One can quantize the system introducing creation and annihilation operators $a_i\equiv e^i_z$ and $a^{\dagger}_i=e^i_{\bar z}$ with canonical commutation relations that just follow from
(\ref{53}). A change in the conformal structure corresponds to a non trivial change of the vacuum
$a_{i} \to \alpha a_i+\beta a_i^\dagger $ with $|\alpha|^2-|\beta|^2=1$ (Bogoliubov transformation). 
In order to analyse the algebra, it will be convenient to introduce the the following definitions
\beq
e_{z}^{+}&\equiv& a_{+}, \qquad 
e_{z}^{-} \equiv a_{-}, \qquad
e_{z}^{3} \equiv b.
\eeq
where $e^{+}= (e^{1}+i e^{2})$. 
Since the metric is real we have that 
$e_{\bar{z}}^{i} = \bar{e}_{z}^{i}$ hence at the quantum level we have
\be
e_{\bar{z}}^{-}= a_{+}^{\dagger}, \qquad 
e_{\bar{z}}^{+} = a_{-}^{\dagger}, \qquad
e_{\bar{z}}^{3} = b^{\dagger}
\ee
The algebra is thus simply a product of three harmonic oscillators
which reads 
\be
[a_{\pm},a^{\dagger}_{\pm}] = 1,\qquad [b,b^{\dagger}]=1
\ee
Given the frame field we can define the fluxes $\Sigma^{i} =\frac12 \epsilon_{ijk} e^{j}_{A}e^{k}_{B} \epsilon^{AB}$.
A straightforward computation  gives 
\beq
\Sigma^{3} &\equiv & a_{\va +}^{\dagger} a_{\va +} - a_{\va -}^{\dagger}a_{\va -}\n \\
\Sigma^{-} &\equiv&  a_{\va +}b^{\dagger} - a_{\va -}^{\dagger}b \n \\
\Sigma^{+} &\equiv&  a_{\va +}^{\dagger}b- a_{\va -}b^{\dagger}\\
\label{sigma}
\eeq
which satisfy the $\SU(2)$ algebra
\be
[\Sigma^{+},\Sigma^{-}]= \Sigma^{3},\quad [\Sigma^{3},\Sigma^{\pm}]=\pm \Sigma^{\pm}.
\ee
with casimir $\Sigma^{i}\Sigma_{i}= \Sigma^{3}{ (\Sigma_{3}{+1}) }+ 2 \Sigma^{-}\Sigma^{+}$.
We also have the metric\footnote{The relationship with the usual real coordinates metric components is
\be\n ds^2=\frac{1}{2}\left[(g_{xx}-g_{yy}-i2g_{xy}) dz^2+cc \right] +2(g_{xx}+g_{yy}) dzd\bar z\ee.} 
\beq
g_{zz} &=& 2a_{\va +}a_{\va -} + b^{2}\n \\
g_{\bar{z}\bar{z}} &=& 2a_{\va +}^{\dagger}a_{\va -}^{\dagger} + b^{\dagger 2}\n \\
g_{z\bar{z}} &=& a_{\va +}^{\dagger}a_{\va +} + a_{\va -}^{\dagger}a_{\va -}  + b^{\dagger}b
\label{metric}
\eeq
which satisfies the algebra
\be
[g_{zz},g_{\bar{z}\bar{z}}]= 4 g_{z\bar{z}},\quad 
[g_{{z}\bar{z}},g_{\bar{z}\bar{z}}]= 2 g_{\bar{z}\bar{z}},\quad 
[g_{{z}\bar{z}},g_{zz}]= -2 g_{z {z}},\quad 
\ee
Note that this algebra is  an $\mathrm{SL}(2,\R)$ algebra
\be\label{algebra}
[g_{+},g_{-}]= -g_{3},\quad [g_{3},g_{\pm}]=\pm 2  g_{\pm}.
\ee
 with
$$g_{3}= g_{z\bar{z}}, \quad g_{+}\equiv \frac {g_{\bar{z}\bar{z}}}{{2}},\quad  g_{-}\equiv \frac {g_{{z}{z}}}{{2}}$$
and 
\be \label{new}
\det(g)=  g_{3}(g_{3}{+}1)-4 g_{+}g_{-} \ee is the casimir of the $\mathrm{SL}(2,\R)$ algebra.
Therefore the canonical commutation relations (\ref{53}) of our initial $12$-dimensional kinematical phase space at each point
is replaced by the ($6$-dimensional) Lie algebra of  $SU(2) \times SL(2,\R)$ in terms of the new fields.
The metric variables encodes the gauge invariant degrees of freedom while the gauge parameters are encoded into the flux  $\Sigma^i$ variables.

\subsection{Diffeomorphism symmetry}\label{fff}
Here we will clarify the geometric interpretation of the $SL(2,\R)$ Lie algebra satisfied by the metric variables.
We will indeed show that the $SL(2,\R)$ transformations can be identified with area preserving transformations of $e_a^i$
which can be seen as an ultra local residue of the group of tangent diffeomorphisms.
The constraint generating tangent diffeomorphisms is
\be
d(v)=\frac{1}{2}\int_{\partial \Sigma} e_i \wedge {\cL_v} e_i,
\ee
where $\cL_v$ denotes the Lie derivative along the vector field $v$ tangent to the boundary. It is direct to verify that
$\{d(v),d(w)\}=d(\cL_v w)$. Using the identity $\cL_v e_i=d(i_v e^i)+ i_v de_i$ one can verify that $d(v)=\tfrac{1}{2} \int_{\partial \Sigma} d((i_v e^i) e_i)+\int_{\partial \Sigma}(i_v e^i) de_i$. The first term vanishes identically due to the fact that $\partial^2 \Sigma=0$.

\begin{figure}[h] \centerline{\hspace{0.5cm} \(
\begin{array}{c}
\psfrag{u}{$u$}\psfrag{d}{$d$}
\psfrag{l}{$l$}\psfrag{r}{$r$}\psfrag{x}{$x$}
\psfrag{y}{$y$}
\includegraphics[width=5cm]{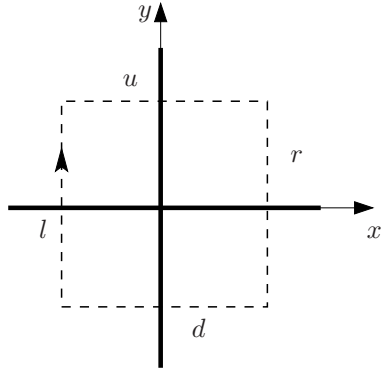}
\end{array}
\)}
\caption{The thick segments represent the paths $L_x$ and $L_y$ used in the regularization of the basic observables used in (\ref{una}). The square oriented path represents the contour $C$ used in (\ref{emi}) defined by four oriented segments $\{u, d, r, l\}$. The diagram  should be thought of as embedded inside a coordinate ball $x^2+y^2\le \epsilon^2$. The regularisation is removed in the limit $\epsilon \to 0$. } \label{figure}
\end{figure}

Let us now assume that the surface $\partial \Sigma$ is decomposed in an union of cells $\partial \Sigma =\cup_i D_i$ with boundaries  $\partial D_i = C_i$.
We can assume for definiteness that each cell $i$ is a square that corresponds to a  lattice cell centered around the vertices of the square lattice introduced in the definition of the basic observables in equations (\ref{una}) and (\ref{otra}).
Let us also assume that inside each cell we impose the Chern-Simons constraints 
$\rd e_i =0$ as a way to express the discretness of our regularisation.
This imposes that the metric is constant within each cell and
this implies that the discrete data determines the value of $e_i$ inside each cell and then on $\partial \Sigma$ (cf. \cite{Freidel:2011ue} for an analog treatment in Loop gravity). Then the diffeomorphism constraint becomes
\be\label{didi}
d(v)=d_{bulk} (v)+\sum_i d_{C_i}(v)
\ee
where the first term generates {\em bulk} diffeomorphisms, which are assumed to vanish, while the $d_{C}(v)'$s are given by 
\ba\label{emi}
\n d_{C}(v)&=&\frac{1}{2}\oint_{C} (i_v e_i) e^i
= \frac{1}{2}\oint_{C} v^a g_{ab} dx^b.
\ea 
In the lattice regularisation one can find an ultra local action of the $d_{C}(v)$ by using paths $C$ as the one depicted in figure \ref{figure}. In that case one finds that
\be\label{cuc}
d_{C}(v)=\delta_1^x g_{xx}+\delta_2^y g_{yy}+(\delta_1^y+\delta_2^x) g_{xy},
\ee 
where $\delta^{a}_1=(v_{u}^a-v_{d}^a)/2$ and $\delta_2^a=(v_{r}^a-v_{l}^a)/2$, and $v_{A}^a$ with $A\in\{u, d, r, l\}$ denotes the value of the vector field at the up, down, right, and left segments respectively defining $C$ as in figure \ref{figure}. We have shown that the action of the $D_C(v)$ corresponds, in our regularisation, to the action of the generators of the $SL(2,\R)$ symmetry that we algebraically deducted from the commutator algebra in (\ref{algebra}).

\subsection{   Representation    }   
We now describe the representations of this metric-flux algebra.
 There is the obvious Fock representation built on top of the vacuum state $|0,0,0\rangle$  anihilated by $b, a_{+},a_{-}$. A general state is denoted by $|n_b,n_+,n_-\rangle$, i.e. the corresponding harmonic oscilator multiparticle states. However, in our case it is more transparent to construct a basis where some of the metric-flux variables are diagonal.
We can describe this algebra in a basis that diagonalises 
$\det(g), \Sigma_{3},g_{3}$.
And we first look for the highest weight states 
which anihilates $g_{-}$.
Such a state  is labelled by a pair of half integers $j,m$ such that $j\pm m \in \N$ and 
can be written
\ba \nonumber
|j,m,{0})&\equiv&{A_{jm}} \sum_{n\in \N_{jm}} (i\sqrt{2})^{n} \frac{(b^{\dagger})^{n}}{{{n!}}} \frac{(a_{+}^{\dagger})^{\frac{j+m-n}{2}}}{{{(\frac{j+m-n}{2})!}}} \frac{(a_{-}^{\dagger})^{\frac{j-m-n}{2}}}{{{(\frac{j-m-n}{2})!}}} |0,0,0\rangle\\
&=& {A_{jm}}{\sum_{n\in \N_{jm}} \frac{(i\sqrt{2})^{n}}{\sqrt {n! (\frac{j+m-n}{2})!(\frac{j-m-n}{2})!}} |n,\frac{j+m-n}{2},\frac{j-m-n}{2}\rangle}
\ea
where the sum is over the ensemble $\N_{jm}$ of all positive integer such that $(j\pm m-n)/2  \in \N$, {and we use $|\ \ )$ instead of $|\ \ \rangle$ to denote the states in the new basis.}

It can be checked that {the previous states form an orthonormal set once  $A_{jm}$ is suitably chosen:
\be\label{sumA}
A_{j,m}^{-2}=\sum_{n\in \N_{jm}} \frac{{2}^{n}}{ {n! (\frac{j+m-n}{2})!(\frac{j-m-n}{2})!}}
\ee
Remarkably this term can be resummed in terms of a simple formula:
\be
A^{2}_{jm} =\frac{ j! (j+m)!(j-m)!}{(2j)!}.
\ee
The proof of this identity can be given by writing the summation formula \ref{sumA}) as an integral:
\be
j! A_{j,m}^{-2}=\int_{-2\pi}^{2\pi} \frac{\rd \phi}{4\pi} ( 2 + e^{i\phi} + e^{-i\phi})^{j} e^{-im \phi} 
=\int_{-\pi}^{\pi} \frac{\rd \psi}{2\pi} (e^{i\psi} + e^{-i\psi})^{2j} e^{-i 2sm \psi}
\ee
where the second equality follows from the change of variable $ \phi = 2\psi$.
We can also check  that $ g_{-}|j,m,0) =0$ while
\be
g_{3}|j,m) = j |j,m,{0}),\qquad \Sigma_{3}|j,m,{0})= m |j,m,{0})
\ee
These states carry a representation of $\SU(2)$ 
given by
\bea
\Sigma^{-} |j,m,{0}) = \frac{-iA_{jm}}{\sqrt{2}A_{jm-1}}(j-m +1) |j, m-1,{0})
= \frac{1}{i\sqrt{2}}\sqrt{(j+m)(j-m+1)} |j, m-1,{0}) \\
 \Sigma^{+} |j,m,{0})
 = \frac{i A_{jm}}{\sqrt{2}A_{jm+1}}(j+m +1) |j, m+1,{0})
 =\frac{i}{\sqrt{2}}\sqrt{(j-m)(j+m+1)} |j, m+1,{0})
\eea
A general state is obtained from these highest weight states by action of $g_{+}$ \be\label{newy}
 g_{+} |j,m,{k}) \equiv {C_{jmk} } \ |j,m,{k+1}) 
\ee
Since $g_3 g_+|j,m,0)=g_+g_3|j,m,0)+g_+|j,m,0)$ on these general states we have
\be\label{we}
g_{3}|j,m,k) = (j+k) |j,m,k),\qquad \Sigma_{3}|j,m,k)= m |j,m,k)
\ee
and the Casimir
\be\label{casi}
\det(g)|j,m,k)=\Sigma_i\Sigma^i |j,m,k)=j(j+1) |j,m,k)
\ee
where
\be
(j,m, k|j',m', k') = \delta_{j,j'} \delta_{m,m'}\delta_{k,k'}\ee
Finally, the operator $g_-g_+$ is also diagonal and plays an important role in the discussion below.
From (\ref{new}) and the commutation relations we get 
\be
g_-g_+=g_+g_-+g_3=\frac{1}{4}(g_3(g_3+1)-\det(g)+4 g_3).
\ee
Now (\ref{casi}) and (\ref{we}) yields 
 \ba\nonumber &&
   g_-g_+ |j,m,k)=\frac{1}{4} ((j+k)(j+k+1)-j(j+1)+4(j+k))|j,m,k), 
 \ea 
 which allows us to compute the coefficients $C_{jmk}$ defined in (\ref{newy}), namely
 \be
 C_{jmk}=\frac{1}{2} {\sqrt{(j+k)(j+k+5)-j(j+1)}}.
 \ee
 
This basically concludes the construction of the representation theory of the geometric observables (\ref{sigma}) and (\ref{metric}). The first surprise is that the condition for the area of the boundary to be finite does not restrict the quantum theory to a finite dimensional Hilbert space. The reason for this is that, even for the zero area eigenstates $j=0$ one has an infinite tower of degenerate excitations $|0,0,k)$ for $k\in \N/2$. In order to recover a finite dimensional subspace defined by a fixed total area one needs to find a way to restricting these $2d$ degenerate geometry quantum number.

\subsection{The geometry of the $k$ quantum number} 

In the absence of external charges, i.e. when $j=0$  (and the local version of the constraint (\ref{45}) is imposed at the quantum level), the only remaining quantum number is $k$. This implies that $k$ is a quantum number associated to the genuine degenerate triads degrees of freedom of the $2+1$ effective theory introduced in Section \ref{eft}.
As mentioned before the presence of these local degrees of freedom is expected from the more general nature of the present boundary conditions which are weaker than those used in the isolated horizon literature. However, such local excitations (encoded in $k$) need to be restricted in some way if we are to recover finite dimensional subspaces that are a key property of the previous treatments. Here we show that there are two natural ways of imposing such restriction. The link with black hole models will be discussed in the conclusion section that follows. 

When $j\not=0$ quantum the  number $k$ admits a geometric interpretation in terms of the metric observables as it follows from
\be\label{goodone}
 g_-g_+ |j,m,k)=\frac{1}{4} [(j+k)(j+k+5)-j(j+1)] |j,m,k),
\ee
which tells that for fixed area eigenvalue (\ref{casi}), or equivalently for fixed $j$, the minimum eigenvalue of $g_-g_+$ is obtained for $k=0$.
Lets us recall here that in conformal coordinates $g_-g_+ = g_{zz}g_{\bz\bz}$ is a measure of the shear deformation of the metric from the diagonal metric.  This means that states picked around the minimal $k$ are (conformally) picked on the fiducial  metric that define our complex structure
\footnote{ Another way of getting a geometric intuition goes as follows: let us make a classical study by writing the triad in our fidutial coordinate system as
\be
e^1=a dx, \ \ \ e^2=b \cos(\phi) dx+b\sin(\phi) dy,\ \ \ e^3=0,
\ee where $e^3=0$ is a partial gauge fixing of the $SU(2)$ symmetry.  A further rotation preserving the condition $e^3=0$ allows us to choose $e^1$ completely ``alined" along $dx$. Now we know that the transformations generated by the metric variables are given by an $SL(2,R)$
of area preserving linear transformations. This means that the $SL(2,\R)$ transformation deform the paralelogram defined by $e^1$ and $e^2$ above without changing its area. If we fix the area to unity we get the condition
\be 1=ab\sin(\phi) \ee
\be
e^1=\frac{a}{\sqrt{2}} (dz+d\bar z), \ \ \ e^2= \frac{b}{\sqrt{2}} (\exp(-i\phi) dz+\exp(i\phi) d\bar z),\ \ \ e^3=0,
\ee
from where we get
\ba\nonumber 
&& g_-=\frac{1}{4}(a^2+b^2 \exp(2i\phi) )\\
&& \nonumber g_+=\frac{1}{4}(a^2+b^2 \exp(-2i\phi) )\\
&& g_3=\frac{1}{2}(a^2+b^2)
\ea
We conclude that the condition $g_-=0$ imply $a^2=b^2$ and $\phi=\pi/2$, from $ab \sin(\phi)=1$ we get $ab=1$ and finally $g_3=1$ or $e^1=dx$, $e^2=dy$, and $e^3=0$. All this is the classical counterpart of the
metric-flux spectral form found above. 
}, namely
\ba
\n \langle g_{z\bar z}\rangle=\sqrt{\det g}=j, \\
\langle g_{zz}\rangle=\langle g_{\bar z\bar z}\rangle=0,
\ea 
and have minimal uncertainties in the off diagonal components that vanish in the large $j$ limit
\be\label{fluc}
\frac{(\Delta g_{xy})^2}{{\det(g)}} \propto \frac{\langle g_-g_{+}\rangle}{\det(g)}=\left.\frac{[(j+k)(j+k+5)-j(j+1)]}{4j(j+1)} \right|_{k=0}=o({1/j})
\ee
The previous semiclassical properties imply that maximum weight states are indeed generalized 
coherent states representing a semiclassical conformally spherical geometry of the boundary. 
The quantum number $k$ is related to (ultra) local diffeomorphims that make the $x$ and $y$ 
directions---canonically chosen by our conformal structure at the starting point---non orthogonal
in the physical metric. Preserving the condition $k=0$ implies the restriction to conformal transformations---diffeomorphisms which preserve the conformal structure at each non trivial ($j\not=0$) puncture.

There is an alternative and equally geometric way of imposing the restriction $k=0$. It corresponds in essence to the
$U(1)$ treatment of \cite{ABK}. The key equations are (\ref{we}). According the algebra (\ref{algebra}) of metric variables, the metric component $g_3$ generates a subgroup $U(1)\subset SL(2,\R)$ corresponding to area preserving diffeomorphisms that can be interpreted as local rotations along a direction normal to the boundary \footnote{In the normal gauge $e^3=0$ we can write $e^1=e^{\phi} (\cos(\theta) dx+\sin(\theta) dy)$
and $e^2=e^{\phi} (-\sin(\theta) dx+\cos(\theta) dy)$
$\sqrt{g}=e^{2 \phi} dx\wedge dy$. The metric component $g_3=g_{z\bar z}=(e^i_xe^i_x+e^i_ye^i_y)/2$ generates 
the transformations \ba
 &&  \n \delta e^A_x=\{g_3,e^A_x\}=e^A_y,\\
&& \n \delta e^A_y=\{g_3,e^A_y\}=-e^A_x.
\ea 
Therefore it is conjugate to the coordinate $\theta$ and generates local rotations of the coordinates around the origin. Notice that one can directly obtain such local differ from the action of $D_C(\partial_\theta)$ as defined in (\ref{cuc}).}. By setting $m=j$ in (\ref{we}) one choses   $SU(2)$ coherent states picked along the internal direction $3$. One can then impose the constraint
\be\label{g3}
g_3-\Sigma_3=0
\ee   
strongly which boils down to setting $k=0$. The previous constraint can be interpreting as aligning the internal direction $3$ with the normal to the normal to the boundary. It links, in this way, the subgroup $U(1)\subset SL(2,\R)$ with the internal subgroup $U(1)\subset SU(2)$. Notice that the vectors $|j,j,0)$ solving the constraint (\ref{g3}) are the only common representation vectors shared by the unitary representations of $SU(2)$ and $SL(2,\R)$ (in the discrete series). 

If no restriction on $k$ is imposed then we have a completely general quantum geometry of the boundary degrees of freedom. The interpretation of $k$ in terms of intrinsic degenerate geometries follows from our analysis of the boundary dynamical system of Section \ref{eft}.

 \section{Conclusion}\label{III}
 
 A simple symplectic structure for the geometry of a $2$-dimensional boundary arises from the canonical formulation of gravity in connection variables. This was previously observed in studies of the isolated horizon boundary condition \cite{AKLR}.
Here we emphasise here its more general validity. 

Starting from such simple symplectic form of the boundary $2$-geometry, expressed in terms of the induced triad field in equation (\ref{sympl}), we have produced a quantisation of the boundary geometry which differs from the one found in the models using a Chern-Simons theory effective treatment.  The main difference consist of the presence of purely degenerate (zero area) point like excitations of the form $|0,0,k\rangle$. Such dissimilarity  should not be surprising as the classical equivalence between description of the boundary geometry presented here and that defined in terms of Chern-Simons theory  is only valid when one assumes the non degeneracy of the boundary geometry (in addition to classical restrictions of symmetry contained in the type I isolated horizon boundary condition)\cite{Ashtekar:2004cn}. The quantisation presented here is therefore more general.     

In order to establish a link with previous formulations one has to supplement our quantisation with an additional  requirement restricting the quantum number $k$ to be equal to zero.
%
This can always be achieved at the classical level by a diffeomorphism. In order to relate our quantisation to the usual treatment we have to impose the diffeomorphism symmetry  at the quantum level. Because the generators of diffeomorphism encoded in the metric components are non commutative it cannot be done strongly. We have discussed here two different ways to proceed. The first possibility is to  require that the averaged complex structure used in the quantisation process matches the one defined by the quantum geometry. 
Such requirement cannot be imposed strongly due to the uncertainty relations but it can be weakly imposed in the semiclassical sense of expectation values and that fluctuations go to zero in the large $j$ limit, equation (\ref{fluc}). This implies the condition $k=0$ is optimal. 
The second possibility is to impose the geometric requirement that the eigenvalues of the generator of $U(1)\subset SL(2,\R)$ area preserving diffeomorphisms coincide with those of the generator of the $U(1)\subset SU(2)$. This condition can be imposed strongly as an operator equation, equation (\ref{g3}). In this second case there is no ambiguity and the restriction sets $k=0$ and $m=j$. The subspace of admissible states at an excited puncture, i.e. $j\not=0$, is one dimensional. This possibility is geometrically very appealing as it links the notions of internal rotations with tangent rotations as defined by the complex structure defining in a way an intrinsically defined normal gauge fixing.
 
 Ultimately  a proper imposition of the diffeomorphism constraints should be investigated. We expect that this will lead to a relationship with conformal field theories in 2d. We leave these appealing aspects for future investigation. 
 The appearance of new degree of freedom associated with diffeomorphisms might provide a concrete example of the kind of non dissipative information reservoir needed in the scenario of unitary black hole evaporation advocated in \cite{Perez:2014xca}.


\begin{thebibliography}{10}

\bibitem{G.:2015sda}
Fernando Barbero and Alejandro Perez.
\newblock {Quantum Geometry and Black Holes}.
\newblock {\em arXiv:1501.02963}, 2015.

\bibitem{BarberoG.:2012ae}
J.~Fernando Barbero~G., Jerzy Lewandowski, and Eduardo~J.S. Villasenor.
\newblock {Quantum isolated horizons and black hole entropy}.
\newblock {\em PoS}, QGQGS2011:023, 2011.

\bibitem{Ashtekar:2004cn}
Abhay Ashtekar and Badri Krishnan.
\newblock {Isolated and dynamical horizons and their applications}.
\newblock {\em Living Rev.Rel.}, 7:10, 2004.

\bibitem{ABK}
Abhay Ashtekar, John Baez, and Kirill Krasnov.
\newblock {Quantum Geometry of Isolated Horizons and Black Hole Entropy}.
\newblock {\em Adv.Theor.Math.Phys.}, 4:1--94, 2000.

\bibitem{Engle:2009vc}
Jonathan Engle, Alejandro Perez, and Karim Noui.
\newblock {Black hole entropy and SU(2) Chern-Simons theory}.
\newblock {\em Phys.Rev.Lett.}, 105:031302, 2010.

\bibitem{Engle:2010kt}
Jonathan Engle, Karim Noui, and Alejandro Perez.
\newblock Black hole entropy from the {SU}(2)-invariant formulation of type {I}
  isolated horizons.
\newblock {\em Phys. Rev. D}, 82, 2010.

\bibitem{Perez:2010pq}
Alejandro Perez and Daniele Pranzetti.
\newblock {Static isolated horizons: SU(2) invariant phase space, quantization,
  and black hole entropy}.
\newblock {\em Entropy}, 13:744--777, 2011.

\bibitem{Bodendorfer:2013sja}
N.~Bodendorfer.
\newblock {Black hole entropy from loop quantum gravity in higher dimensions}.
\newblock {\em Phys.Lett.}, B726:887--891, 2013.

\bibitem{Freidel:2014qya}
Laurent Freidel and Yuki Yokokura.
\newblock {Non-equilibrium thermodynamics of gravitational screens}.
\newblock 2014.

\bibitem{Barbero:1994ap}
J.~Fernando Barbero~G.
\newblock {Real Ashtekar variables for Lorentzian signature space times}.
\newblock {\em Phys.Rev.}, D51:5507--5510, 1995.

\bibitem{Immirzi:1996dr}
Giorgio Immirzi.
\newblock {Quantum gravity and Regge calculus}.
\newblock {\em Nucl.Phys.Proc.Suppl.}, 57:65--72, 1997.

\bibitem{Ashtekar:2004nd}
Abhay Ashtekar, Jonathan Engle, and Chris Van Den~Broeck.
\newblock {Quantum horizons and black hole entropy: Inclusion of distortion and
  rotation}.
\newblock {\em Class.Quant.Grav.}, 22:L27--L34.

\bibitem{Beetle:2010rd}
Christopher Beetle and Jonathan Engle.
\newblock {Generic isolated horizons in loop quantum gravity}.
\newblock {\em Class.Quant.Grav.}, 27:235024, 2010.

\bibitem{Frodden:2012en}
Ernesto Frodden, Alejandro Perez, Daniele Pranzetti, and Christian Röken.
\newblock {Modelling black holes with angular momentum in loop quantum
  gravity}.
\newblock {\em Gen.Rel.Grav.}, 46(12):1828, 2014.

\bibitem{Freidel:2011ue}
Laurent Freidel, Marc Geiller, and Jonathan Ziprick.
\newblock {Continuous formulation of the Loop Quantum Gravity phase space}.
\newblock {\em Class.Quant.Grav.}, 30:085013, 2013.

\bibitem{AKLR}
Abhay Ashtekar and Badri Krishnan.
\newblock Isolated and dynamical horizons and their applications.
\newblock {\em Living Rev. Relativity}, 7, 2004.

\bibitem{Perez:2014xca}
Alejandro Perez.
\newblock {No firewalls in quantum gravity: the role of discreteness of quantum
  geometry in resolving the information loss paradox}.
\newblock {\em Class.Quant.Grav.}, 32(8):084001, 2015.

\end{thebibliography}
\end{document}